\documentclass[11pt]{article}

\def\boxit#1{\vbox{\hrule\hbox{\vrule\kern6pt
          \vbox{\kern6pt#1\kern6pt}\kern6pt\vrule}\hrule}}

\usepackage{xr-hyper}  

\usepackage[hyperindex=true,pdftitle={},
pdfauthor={St\'ephane Guerrier},colorlinks=TRUE,
pagebackref=false,citecolor=blue, urlcolor = blue, plainpages=false,
pdfpagelabels]{hyperref}

\usepackage[round]{natbib}  

\usepackage[font=small, labelfont=sc, textfont = it]{caption}
\usepackage{bm}
\usepackage{amsthm} 
\usepackage{color}
\usepackage{amssymb}
\usepackage{amsbsy}
\usepackage{amsfonts}
\usepackage{amsmath}
\usepackage{color}
\usepackage{graphics}
\usepackage[pdftex]{graphicx}
\usepackage{setspace} 
\usepackage{rotating}

\usepackage[sc]{titlesec}

\usepackage{placeins}

\usepackage{footnote}

\usepackage{mathtools}

\usepackage{nicefrac}

\usepackage[top=1.5in, bottom=1.5in, left=1.4in, right=1.4in]{geometry}
\usepackage[protrusion=true,expansion=true]{microtype}	
\usepackage{booktabs}
\usepackage{longtable}

\DeclareMathOperator*{\argmin}{argmin}

\def\boxit#1{\vbox{\hrule\hbox{\vrule\kern3pt
          \vbox{\kern3pt#1\kern3pt}\kern3pt\vrule}\hrule}}

\definecolor{pinegreen}{rgb}{0.0, 0.47, 0.44}		
\newtheoremstyle{mytheoremstyle} 
    {0.3cm}                      
    {0cm}                        
    {\itshape}                   
    {}                           
    {\scshape}                   
    {: }                          
    {0em}                       
    {}  

\theoremstyle{mytheoremstyle}

\newtheoremstyle{myExampleRemarkstyle} 
    {0.3cm}                    
    {0cm}                           
    {\itshape}                   
    {}                           
    {\scshape}                   
    {: }                          
    {0em}                       
    {}  

\theoremstyle{myExampleRemarkstyle}

\newtheoremstyle{simuStyle}
{0.3cm} 
{0cm} 
{} 
{} 
{\bfseries} 
{.} 
{0em} 
{} 

\theoremstyle{simuStyle}

\newtheoremstyle{stratStyle}
{0.3cm} 
{0cm} 
{} 
{} 
{\scshape} 
{: } 
{0em} 
{} 

\theoremstyle{stratStyle}

\DeclareSymbolFont{lettersA}{U}{txmia}{m}{it}
\DeclareMathSymbol{\real}{\mathord}{lettersA}{"92}
\DeclareMathSymbol{\field}{\mathord}{lettersA}{"83}

\makeatletter							
\def\printtitle{%
    {\centering \@title\par}}
\makeatother									

\makeatletter							
\def\printauthor{%
    {\centering \large \@author}}				
\makeatother

\newlength{\tpheight}
\newlength{\txtheight}
\newlength{\tpwidth}
\newlength{\txtwidth}

\def\E{\mathbb{E}}
\def\wh{\widehat}  
\def\X{{\bf X}}
\def\bb{\boldsymbol{\beta}}
\def\trans{{^{\rm T}}}

\begin{document}


\begin{center}
\Huge{\textsc{A Paradigmatic Regression Algorithm for Gene Selection Problems}}
  \vspace{0.75cm}\\
  \small{\textsc{St\'ephane Guerrier}$^{\ddag \,\, \star}$, \textsc{Nabil Mili}$^{\dag \,\, \star}$, \textsc{Roberto Molinari}$^{\dag}$,} \\
\small{\textsc{Samuel Orso}$^{\dag}$, \textsc{Marco Avella-Medina}$^{\dag}$ \& \textsc{Yanyuan Ma}$^\S$} \vspace{0.75cm}\\
  
  $^{\ddag}$\small{\textsc{Department of Statistics}}\\
  \small{\textsc{University of Illinois at Urbana-Champaign, USA}}\\
  \small{\textsc{Email:}}	\href{mailto:stephane@illinois.edu}{stephane@illinois.edu} \vspace{0.25cm}\\

  $^{\dag}$\small{\textsc{Research Center for Statistics}}\\
  \small{\textsc{Geneva School of Economics and Management}}\\
  \small{\textsc{University of Geneva, Switzerland}}\\
  \vspace{0.25cm}

  $^{\S}$\small{\textsc{Department of Statistics}}\\
  \small{\textsc{University of South Carolina, USA}}\\
\end{center}

\footnotetext{$^{\star}$ The first two authors are Joint First Authors.} 


\begin{abstract}
\textsc{\textbf{Motivation:}}
Gene selection has become a common 
task in most gene expression studies. The objective of such research is 
often to identify the smallest possible set of genes that can still 
achieve good predictive performance. The problem of assigning tumours to 
a known class is a particularly important example that has 
received considerable attention in the last ten years. Many of the 
classification methods proposed recently require some form of
dimension-reduction of the problem. These methods provide a single
model as an output and, in most cases, rely
on the likelihood function in order to achieve variable selection.  \\

\textsc{\textbf{Results:}}
We propose a prediction-based 
objective function that can be tailored to the requirements of 
practitioners and can be used to assess and interpret a given 
problem. The direct optimization of such a function can be very difficult because
the problem is potentially discontinuous and nonconvex. We therefore 
propose a general procedure for variable selection that resembles
importance sampling to explore the feature space.
Our proposal compares favorably with competing 
alternatives when applied to two cancer data sets in that smaller 
models are obtained for better or at least comparable classification 
errors. Furthermore by providing a set of selected models instead of a 
single one, we construct a network of possible models for a target
prediction accuracy level.


\textsc{\textbf{Contact:}} 
\href{stephane@illinois.edu}{stephane@illinois.edu}

\end{abstract}

\section{Introduction} 

Gene selection has become a common task in most gene expression
studies. The goal of this research is often to identify the
smallest possible set of genes that can still achieve good predictive
performance \citep{diaz2006gene}.  The problem of assigning tumours to
a known class is an example that is of particular importance and has
received considerable attention in the last ten years. Conventional class
prediction methods of leukemia or other cancers are in general based
on microscopical examination of stained tissue specimens. However,
such methods require highly trained specialists and are subjective
\citep{tibshirani2002diagnosis}.  

To avoid these drawbacks, many automatic classification methods have
been proposed recently. These methods have the advantage of being
objective and have 
improved the correct classification rate in various cases.
Among the different methodologies brought forward in this context we can find those proposed by
\citet{tibshirani2002diagnosis}, \citet{dudoit2002comparison},
\citet{zhu2004classification}, \citet{zou2005regularization}. See also
\citet{diaz2006gene} and the references therein for other approaches.  

Nonetheless, many of these methods do not necessarily respond to the
needs of practitioners and researchers when they approach the gene
selection process. First of all, many of them have to rely on some
form of size reduction and often require a subjective input to
determine the dimension of the problem. Also, many of these methods
often provide a single model as an output whereas genes interact
inside biological systems and can be interchangeable in explaining a
specific response. The idea of interchangeability of genes in
explaining responses appears for instance in
\citet{kristensen2012integrated}. These authors use the PARADIGM
algorithm of \cite{vaskeetal2010} to combine  mRNA expression and DNA
copy number in order to construct clusters of patients that provide
the best predictive value. The resulting clusters can be seen as being
characterized by different significantly expressed genes.

Another issue of most existing gene selection methods is their
reliance on the likelihood function, or a penalized version of it, as
a means to develop a selection criterion. However, the likelihood
function may not necessarily be the quantity that users are interested
in as they may want to target some other kind of loss function such as, for example, the classification error.  
Of course, maximizing the likelihood function is not typically the
same as minimizing a particular loss function. 
Moreover, adapting these methods to handle missing or contaminated
data is not straightforward. This has limited the applicability and
reliability of these methods in many practical cases. 

To eliminate the limitations of the gene selection procedures
described above, this paper proposes an objective function for
out-of-sample predictions that can be tailored to the requirements of
practitioners and researchers. This is achieved by enabling them to
select a criterion according to which they would like to assess and/or
interpret a given problem. However, the optimization of such a
criterion function is typically not an easy task since the function
can be discontinuous, non-convex and would require computationally
intensive techniques. To tackle this issue, we propose a solution
using a different approach based on a procedure that resembles
\textit{importance sampling}. This new approach provides a general and
flexible framework for gene selection as well as for other model
selection problems. 

The advantages of this proposal are multiple:

\begin{itemize}

    \item \textbf{Flexibility}: It allows the users to specify a
      criterion that can be tailored to the specific problem
      setting. It is able to handle different kinds of responses,
      problems of missing and contaminated data, multicollinearity,
      etc. 

    \item \textbf{Prediction Power}: The result of the procedure
      is a set of models with high predictive power with respect to
      the specified criterion. It is especially suitable in selecting
      genes and models to achieve accurate predictions. 

    \item \textbf{Dimension-reduction}: It can provide an assessment
      of the dimension of the problem because it greatly reduces the
      number of necessary covariates and eases the interpretation
      without requiring any preliminary size reduction. 

    \item \textbf{Network-building}: With the reduced model size, it
      preserves the capacity to build gene-networks to provide a more
      general view of the potential paradigmatic structures of the
      genetic information. 
\end{itemize}
 
This last aspect is of great interest for gene selection since this
list can provide insight into the complex mechanisms behind different
biological phenomena. Different cases, some of which can be found in
Section \ref{results}, indicate that this method appears to outperform
other methods in terms of criteria minimization while, at the same
time, selects models of considerably smaller dimension which allow
improved interpretation of the results. The set of selected models can
naturally be viewed as a network of possible structures of genetic
information. We call this a paradigmatic network. In Section
\ref{results} we give an example of a graphical representation of
such networks based on the analysis of one of two cancer data sets which are discussed therein.  

In this paper we first describe and formalize the proposed
approach within the model selection statistical framework  in Section
\ref{approach}. In Section \ref{methods} we   illustrate the
techniques and algorithms used to address the criterion minimization
problem highlighted in Section \ref{approach}. The performance  of our
approach is then illustrated on two data sets concerning leukemia classification
\citep{golub1999molecular} and breast cancer classification
\citep{chin2006genomic} in Section \ref{results}. We
conclude the paper in Section \ref{conclusion} by summarizing the
benefits of the new approach and providing an outlook on other
potential applications that can benefit from this methodology. 

%

\section{Approach}
\label{approach}

To introduce the proposed method, let us first define some notation
which will be used throughout this paper: 
\begin{enumerate}
	\item Let $\mathcal{J}_f = \left\{1,2,...,p\right\}$ be the set of indices for  $p$ potential covariates included in the $n \times p$ matrix $\X$. We allow $\X$ to include a vector of 1s. 
	\item Let $\mathcal{J} = \mathcal{P}(\mathcal{J}_f) \setminus \emptyset$, $|\mathcal{J}| = 2^p-1$, be the power set including all possible models that can be constructed with           the $p$ covariates excluding the empty set.
	\item Let $\jmath \in \mathcal{J}$ be a model belonging to the above mentioned power set. 
	\item Let $\bb^{\jmath} \in {\mathbb R}^{p}$ be the parameter vector for model $\jmath$, i.e.  
	\begin{equation*}
	 \bb^{\jmath}_k=
	 \left\{
	\begin{array}{ll}
		\bb_k  & \mbox{if } \ k \in \jmath \\
		0 & \mbox{if } \ k \not\in \jmath 
	\end{array}
\right. \nonumber
	\end{equation*}	
where $\bb_k^{\jmath}$, $\bb_k$ are respectively the $k$th element of
$\bb^{\jmath}$ and $\bb$.  

\end{enumerate}

Keeping this notation in mind, for a given model ${\jmath} \in \mathcal{J}$ we have
	\begin{equation}
	Y = g(\X,\bb^{\jmath}) + \varepsilon,
	\label{eq:model}
	\end{equation}	
	where $g(\cdot,\cdot)$ is a link function known up to the parameter
        vector $\bb^{\jmath}\in\mathbb{R}^{p}$ and $\varepsilon$ is a
        mean zero random error.  
%
Models of the form (\ref{eq:model}) are very general and include all
parametric models and a large class of semiparametric models when
$g(\cdot,\cdot)$ is not completely known or the distribution of
$\varepsilon$ is not specified. A few examples of model (\ref{eq:model}) are given in Appendix \ref{sec:example}.
%
%

We assume that for a fixed $\jmath$,  based on a specific choice for model (\ref{eq:model})
with corresponding parameter vector $\bb^{\jmath}$ and given a new covariate vector
$\X_0$, the user can construct a prediction
$\wh{Y}(\X_0,\bb^{\jmath})$. To assess the quality of this prediction we assume that we have a divergence measure available which we denote as $D\{
\wh{Y}(\X_0,\bb^{\jmath}), Y_0\}$. The only requirement imposed on the
divergence measure is that it satisfies the property of positiveness,
i.e.
	\begin{equation*}
		\begin{aligned}
			D(u,v) &> 0 \;\; \text{for} \;\; u \neq v \\
			D(u,v) &= 0 \;\; \text{for} \;\; u = v.
		\end{aligned}
	\end{equation*}
With this property being respected, the divergence measure can arbitrarily be specified by the
user according to the interest in the problem. Examples of such divergence measures include the
$L_1$ loss function 
\begin{equation*}
D\{\wh{Y}(\X_0,\bb^{\jmath}), Y_0\}=|\wh{Y}(\X_0,\bb^{\jmath})-Y_0|
\end{equation*}
or an asymmetric classification error
\begin{equation*}
	\begin{aligned}
		D\{\wh{Y}(\X_0,\bb^{\jmath}), Y_0\} = & I\{\wh{Y}(\X_0,\bb^{\jmath})=1, Y_0= 0\}w_1 \\
		+ & I\{\wh{Y}(\X_0,\bb^{\jmath}) =0, Y_0= 1\} w_2 .
	\end{aligned}
\end{equation*}
where $w_1, w_2\geq0$. The latter is for a Bernoulli response and is typically
an interesting divergence measure when asymmetric classification
errors have to be considered. Indeed, in most clinical situations, the
consequences of classification errors are not equivalent  
with respect to the direction of the misclassification. For instance,
the prognosis and the treatment of Estrogen Receptor (ER) positive
Breast Cancers (BC) are quite different from thoe of ER negative
ones. Indeed, if a patient with ER negative is
  treated with therapies designed for patients with ER positive, the
  consequence is much more severe than if this were done the other way round because of
  the excessive toxicities and potentially severe side effects.
It therefore makes sense to give  different values to $w_1$ and $w_2$. By defining
 $w_1 > w_2$ we would take these risks into account, where $w_1$ would be the weight for a misclassification from
 ER negative to ER positive BC and $w_2$ for the opposite
 direction. Weight values can be modulated according to the current
 medical knowledge and the clinical intuition of the physicians.


Considering this divergence measure $D(\cdot,\cdot)$, we are consequently interested in finding the best models within the general class given in \eqref{eq:model}. To do so, we would ideally aim at solving the following risk minimization problem :
\begin{equation}
 \wh \bb^{\jmath} \in \mathcal{B} \equiv \underset{{\jmath} \in \mathcal{J}}{\argmin} \, \underset{\bb^{\jmath}}{\argmin} \; \mathbb{E}_0\left[D\left\{\wh{Y}(\X_0,\bb^{\jmath}),Y_0\right\}\right],
	\label{eq:opt.prob}
\end{equation}
where 
 $\mathbb{E}_0$ denotes the expectation on the new observation $(Y_0, \X_0)$.
Let $\jmath_0$ denote the models with the smallest cardinality among
all $\wh\bb^\jmath\in\mathcal{B}$. 
Note that there could be more than one model with the same prediction
property and of the same size, hence $\jmath_0$
could contain more than one model.
Let us define the models corresponding to $\jmath_0$ as the
 ``true'' models. Thus,  our ``true'' models are
 essentially the most parsimonious models that minimize the expected prediction error.

The optimization problem in (\ref{eq:opt.prob}) is typically very
difficult to solve.  First of all, supposing we do not consider interaction terms, the outer minimization would require to compare a total of
$2^p-1$ results, each a result of the inner minimization problem. In
addition, each of the $2^p-1$ inner minimization problems is also very
hard to solve,  even if the risk
$\mathbb{E}_0[D\{\wh{Y}(\X_0,\bb^\jmath),Y_0\}]$ were a known function of
$\bb^\jmath$. Indeed, the inner minimization problem is in general
non-convex and could be combinatorial, implying that the minimizer might not be
unique. For example, when $D(\cdot,\cdot)$ is the classification
error, this problem is combinatorial by nature. 
In practice, the computational challenge is even greater because 
the risk function $\mathbb{E}_0[D\{\wh{Y}(\X_0,\bb^\jmath ),Y_0\}]$ is a
function of $\bb^J$ without explicit form
and needs to be approximated.

We propose to estimate $\mathbb{E}_0[D\{\wh{Y}(\X_0,\bb^{\jmath}),Y_0\}]$ 
via an $m$-fold cross-validation (typically $m = 10$) repeated $K$ times.
More specifically, for a sample of size $n$, we repeat the following procedure $K$ times. 
At the $k$th repetition, we randomly select $\lfloor n/m \rfloor$
observations to form a ``test'' data set, subindexed $i=1, \dots, \lfloor
n/m \rfloor$ and superindexed $k=1, \dots, K$, i.e. $(\X_i^k, Y_i^k)$,
then the estimated risk is 
\begin{equation}
\wh \E_0\left[D\left\{\wh{Y}(\X_0,\bb^{\jmath}) ,Y_0\right\}\right]
= \frac{1}{\lfloor n/m\rfloor K} \sum_{k=1}^K \sum_{i=1}^{\lfloor n/m \rfloor} D\{\wh{Y}(\X_i^k,\bb^\jmath),Y_i^k\}.
	\label{eq:crit.estim}
\end{equation}
The reason we only use $\lfloor n/m \rfloor$ observations out of the whole
data set will become clear further on.
Having approximated the expectation $\E_0$, the minimization problem
in (\ref{eq:opt.prob}) becomes
\begin{equation}\label{eq:opt.prob.sample}
 \underset{\jmath \in \mathcal{J}}{\argmin} \,\,
 \underset{\bb^{\jmath}}{\argmin} \,\,
\wh \E_0\left[D\left\{\widehat{Y}(\X_0,\bb^{\jmath}),Y_0\right\}\right].
\end{equation}
Despite the above approximation, the minimization problem remains
complicated for the reasons mentioned earlier.
Thus, we further eliminate the inner minimization problem in
(\ref{eq:opt.prob.sample}) by inserting an estimator $\wh\bb^\jmath$
obtained independently from the minimization procedure. More specifically,
we assume that an estimator of $\bb^{\jmath}$, say
$\wh\bb^{\jmath,k}$, is available based on model (\ref{eq:model}) and ``training'' 
observations $(\X_{\lfloor n/m \rfloor +1}^k, Y_{\lfloor n/m \rfloor
  +1}^k), \dots, (\X_n^k, Y_n^k)$ (i.e. those observations excluded from the above mentioned ``test'' data sets). This estimator can be any available estimator, for example, the maximum likelihood estimator (MLE), a
moment based estimator, or a quantile regression based estimator, etc.
We then replace the inner minimization in (\ref{eq:opt.prob.sample})
directly with the approximate expectation evaluated at $\wh\bb^{\jmath,k}$'s and 
simplify (\ref{eq:opt.prob.sample}) to 
	\begin{equation*}	
\underset{\jmath \in \mathcal{J}}{\argmin} \;  \frac{1}{\lfloor n/m \rfloor K} \sum_{k=1}^K \sum_{i=1}^{\lfloor n/m \rfloor} D\left\{\wh Y(\X_i^k,\wh{\bb}^{\jmath,k}),Y_{i}^k\right\}.
	\label{eq:approx.prob} 
	\end{equation*}	
The intuition of replacing the inner minimization in
(\ref{eq:opt.prob.sample}) with a sample average evaluated at an
arbitrary estimator 
is due to the fact that this estimator, under a fixed ``true'' model and regardless of whether this
estimator is a standard MLE or a minimizer of the divergence measure
$D$, is an approximation to the ``true'' parameter. This means that, consequently, different
estimators are ``close'' to each other.  As a consequence,
$ ( \lfloor n/m \rfloor K )^{-1}\sum_{k=1}^K \sum_{i=1}^{ \lfloor n/m \rfloor} D\left\{\wh Y(\X_i^k,\wh{\bb}^{\jmath,k}),Y_{i}^k\right\}$ is a close approximation to 
$\min_{\bb^\jmath}\mathbb{E}_0\left[D\left\{\wh{Y}(\X_0,\bb^{\jmath}),Y_0\right\}\right]$.

We now have an optimization problem in (\ref{eq:approx.prob}) which
requires a comparison of $2^p-1$ values and is much easier to
solve. To further reduce the number of comparisons, the following section describes some procedures and algorithms allowing to solve this problem in a more efficient manner.

\section{Heuristic procedure}
\label{methods}
To solve the optimization problem in
(\ref{eq:approx.prob}), we propose an approach designed to have the
following three features: 
\begin{enumerate}
    \item Identify a \textbf{set of models} that carry large predictive power
      instead of a single ``best'' model;
    \item Find this set of models within a \textbf{reasonable time}, without
      having to explore all possible models;
    \item This set achieves \textbf{sparsity}, i.e. most of
      the parameters in $\bb$ will be fixed at zero in each of the
      models in the set. 
\end{enumerate}

Note that the last feature above reflects the belief that most of the
covariates are irrelevant for the problem under
consideration and should be excluded. Indeed, our method is designed
to work effectively if such a sparsity assumption holds, putting it on
the same level of almost all 
variable selection procedures in the literature. Moreover, we require the method to have the first feature in order to increase
flexibility in terms of interpretation. Indeed, in many domains such
as gene selection, for example, the aim may not be to find a single
model but a set of variables (genes) that can be inserted in a
paradigmatic structure to better understand the contribution of each
of them via their interactions.

Given this goal, assume that we have at our disposal an estimate of the measure of
interest $D(\cdot,\cdot)$  for all possible $2^p-1$ models. In this
case, our interest would be to select a set of
``best'' models by simply keeping the set of models that have a low
discrepancy measure $D(\cdot,\cdot)$. It is of course unrealistic to obtain a discrepancy measure for all
models in most practical cases because this would require a
considerable amount of time for computation. Therefore, in order to achieve the second feature, instead of examining all possible models, we
can randomly sample covariates from $\mathcal{J}$. The random sampling 
needs to be carefully devised because 
in practice, for example in gene selection problems, the number of
covariates $p$ can easily reach thousands or tens of thousands (see 
examples in Section \ref{results}, where $p=7,129$ and
$p=22,215$ respectively).
 In such situations, $2^p-1$ is an extremely large number and the probability
 of randomly sampling a ``good'' set of variables from the $2^p-1$ variables is very small.
Using the sparsity property of the problem, we propose to start with
the set of variables $\mathcal{M}_0$ (typically an empty set)
and increase the model
complexity stepwise. Throughout this procedure, we ensure that at step
$k$, the more promising covariates based on the evaluation at step $k-1$
are given higher probabilities of being randomly drawn. The last idea
is in the spirit of ``importance sampling'' in the sense that
covariates with more importance based on the previous step are
``encouraged'' to be selected in the current step. Note that by
construction we achieve sparsity if we stop the stepwise search at
models of size $d_{\text{max}}\ll p$. 

More formally, let us first define the set of all possible models of size
$d$ as 
\begin{equation*}
    \mathcal{S}_d = \{ ( i_1,\ldots,i_d ) \; | \; i_1,\ldots,i_d \in \mathcal{J}_f; \; i_1<\ldots<i_d \}.
\end{equation*}
We then define the set of promising models, $\mathcal{S}_d^*$,  as the
ones with an estimated 
out-of-sample divergence measure $D(\cdot,\cdot)$ below a certain
estimated $\alpha$-quantile. The
value of $\alpha$ is user-defined depending on the problem at hand,
and is typically a small value such as $\alpha=1\%$. The formal definition of this set would then be
\begin{equation*}
    \mathcal{S}_d^* = \{\jmath \; | \; \jmath \in \mathcal{S}_d \; ; \; \wh D_\jmath \leq \wh{q}_d(\alpha)\},
\end{equation*}
where \begin{equation}\label{eq:crit:algo}
    \wh D_\jmath \equiv \frac{1}{\lfloor n/m \rfloor K} \sum_{k=1}^K
    \sum_{i=1}^{\lfloor n/m \rfloor} D\{\wh Y(\X_i^k,\wh{\bb}^{\jmath,k}),Y_i^k\},
\end{equation}
and $\wh{q}_d(\alpha)$ is the $\alpha$-quantile of the $\wh D_\jmath \,\, (\jmath\in\mathcal{S}_d)$ values issued from the $B$ randomly selected models.
Finally, we define the set of indices of covariates that are in $\mathcal{S}^*_d$ as 
	\begin{equation*}	\label{eq:formsetI}
    \mathcal{I}^*_d = \{i \; | \; i\in \jmath,  \; \jmath\in \mathcal{S}^*_d\}
	\end{equation*}	
whose complement we define as $\mathcal{I}^c_{d}$ (i.e. all those covariates that are not included in $\mathcal{I}^*_d$).


With this approach in mind and using the above notations, to start the
procedure we assume that we have $p$ variables from which to select.

\begin{enumerate}
	\item[A.] \emph{Initial Step:} We start by adding the number of variables $d = 1$ to our
          initial variable set $\mathcal{M}_0$ 
          with the goal of finally obtaining the set
          $\mathcal{I}^*_1$.
	\begin{enumerate}
		\item[1.] Construct the $p$ possible one variable
                  models by augmenting $\mathcal{M}_0$ with each of the $p$ available variables.
		\item[2.] Compute $\widehat{D}$ for every model obtained in Step A.1.
		\item[3.] From Steps A.1 and A.2, construct the set
                  $\mathcal{I}^*_1$ using (\ref{eq:formsetI}). Go to
                  Step B and let $d = 2$.
	\end{enumerate}
	\item[B.] \emph{General Step:} We define here the general
          procedure to construct $\mathcal{I}^*_d$ for $2 \leq d \leq
          d_{\max}$.
	\begin{enumerate}
		\item[1.] Augment $\mathcal{M}_0$ with $d$ variables as follows:
		\begin{enumerate}
			\item[(i)] Randomly select a set, either  set
                          $\mathcal{I}^*_{d-1}$ with probability $\pi$
                          or its complement $\mathcal{I}^c_{d-1}$ with
                          probability $1 - \pi$. 
			\item[(ii)] Select one variable uniformly at
                          random and without replacement from the set
                          chosen in Step (i) and add this variable to
                          $\mathcal{M}_0$. 
			\item[(iii)] Repeat Steps (i) and (ii) until
                          $d$ variables are added to $\mathcal{M}_0$. 
		\end{enumerate}
		\item[2.] Construct a model of dimension $d$ using the
                  $d$ variables selected in Step B.1. Repeat Step B.1
                  $B$ times to construct $B$ such models. 
		\item[3.]  From Steps B.1 and B.2, construct the set
                  $\mathcal{I}^*_d$ according to
                  (\ref{eq:formsetI}). If $d < d_{\max}$, go to Step B
                  and let $d = d+1$, otherwise exit algorithm. 
	\end{enumerate}
\end{enumerate}

\subsection{Discussion}

Some ideas that our approach and our algorithm use can be found in the literature. We relegate a short discussion of these to Appendix \ref{sec:related}. Here we will discuss instead some practical issues arising when implementing our algorithm. 

\subsubsection{Choice of algorithm inputs:}
The parameters $d_{\text{max}}$, $B$, $\alpha$ and $\pi$ of the above
algorithm are to be fixed by the user. As mentioned earlier
$d_{\text{max}}$ represents a reasonable upper bound for the model
dimension which is constrained to $d_{\text{max}}\leq l$, where $l$
depends on the limitations of the estimation method and is commonly
the sample size $n$. As for the parameter $B$, a larger value is
always preferable to better explore the covariate space. However, a larger $B$ implies heavier
computations, hence
a rule of thumb that could be used is to choose this parameter such that $p \leq B \leq \binom{p}{2}$. As mentioned earlier, the parameter $\alpha$ should define a small quantile, typically 1\%. Finally, $\pi$ determines to what extent the user assigns importance to the variables selected at the previous step. Given that $d_{\text{max}}\ll p$ and $\alpha$ is small, we will typically have that $|\mathcal{I}^*_{d-1}|<|\mathcal{I}^c_{d-1}|$. In this setting, a choice of $\pi=.5$ for example would deliver a higher probability for the variables in $\mathcal{I}^*_{d-1}$ to be included in $\mathcal{I}^*_{d}$. All other parameters being equal, increasing the value of $\pi$ would decrease the probability of choosing a variable in $\mathcal{I}^c_{d-1}$ and vice versa. Moreover, we discuss in Appendix \ref{sec:ada} how the proposed algorithm can be adjusted to situations where $p$ is either small or very large.

As a final note, it is also possible for the initial model
$\mathcal{M}_0$ to already contain a set of $p_0$ covariates which the
user considers to be essential for the final output. In this case, the
procedure described above would remain exactly the same since the
procedure would simply select from the $p$ covariates which are not in
the user-defined set and the final model dimension would simply be
$p_0 + d$.  

\subsubsection{Algorithm output:}
\label{algo:out}

Once the algorithm is implemented, the user obtains an out-of-sample
discrepancy measure for all evaluated models. The final goal is then
to find a subset of models of dimension ${d^*}$ that in some way
minimize the considered discrepancy. A possible solution would be to
select the set of models $\mathcal{S}^*_{d^*}$ such that $d^* = \min_{
  \in\{1,\dots,d_{\text{max}}\}} \; {q}_{d}(\alpha)$. However, the
quantity ${q}_{d}(\alpha)$ is unknown and replaced by its estimator
$\wh{q}_{d}(\alpha)$. Due to this, a solution that might be more
appropriate would be to consider a testing procedure to obtain $d^*$
taking into account the variability of $\wh{q}_{d}(\alpha)$. For
example, we could find the dimension $d^*$ such that we cannot reject
the hypothesis that $\wh{q}_{d^*}(\alpha) =
\wh{q}_{d^*+1}(\alpha)$. Thus we sequentially test whether
$\wh{q}_{j+1}$ is smaller than $\wh{q}_{j}$ for
$j=1,\dots,d_{\text{max}}$. As long as the difference is significant
we increment $j$ by one unit, otherwise the minimum is reached and
$d^*=j$. A more detailed discussion on the type of tests that can be used for this purpose is presented in Appendix \ref{sec:test}.

\FloatBarrier
\section{Case Studies}
\label{results}

    
In this section we provide an example of how the methodology
proposed in this paper selects and groups genes to explain, describe
and predict specific outcomes. 
We focus on the data-set (hereinafter \textit{leukemia}) which collects
information on Acute Myeloid Leukemia (AML) and Acute Lymphoblastic
Leukemia (ALL) and is frequently used as an example for gene selection
procedures. Indeed, \citet{golub1999molecular} were among the first to
use this data to propose a gene selection procedure which was then
followed up by other proposals that used the same data to compare
their performance. We will use this data-set to underline
the features and advantages of the proposed method. A second
data-set concerning the research on
breast cancer (presented in \citet{chin2006genomic}) is analysed in Appendix \ref{sec:breast} to show the outputs of the proposed method from another example.


The analysis of these data-sets focuses both on the advantages of the
proposed methodology and the biological interpretation of the
outcomes. One of the goals of our method is to help
decipher the complexity of biological systems. We
will take on an overly simplified view of the cellular processes
in which we will assume that one biomarker maps to only one gene that
in turn has only one function. Although this assumption is not
realistic, it allows us to give a straightforward
interpretation of the selected models or ``networks'' which can
therefore provide an approximate first insight into the relationships
between variables and biomarkers (as well as between the
biomarkers themselves). We clarify that we
do not claim any causal nature in the conclusions we present in these
analyses but we believe that the selected covariates can eventually be
strongly linked to other covariates that may have a more obvious and
direct interpretation for the problem at hand.
 Finally, the data-set 
has binary outcomes (as does the data-set in Appendix \ref{sec:breast}), hence we will make use of
the Classification Error (CE) as a measure of prediction performance
and we will not assign weights to a given prediction
error. This means that misclassification errors are given the same
weight, in the sense that a false positive prediction (e.g. predicted
``presence'' when the truth is ``absence'') is considered as
undesirable as a false negative prediction. However, our method can
consider also divergence measures based on unequal weights as
highlighted in Section \ref{approach}. 

\subsection{Acute Leukemia}
\label{golub} 

\subsubsection{Statistical analysis}

\citet{golub1999molecular} were among the first
to propose an automatic selection method for cancer classification and
demonstrated the advantages of using such a method. One of the main
applications of their method was on the \textit{leukemia} data-set in
which information regarding 72 patients is included, namely their type
of leukemia (25 patients with AML and 47 patients with ALL) and 7,129
gene expressions used as explanatory variables to distinguish between
two types of leukemia. As explained in \citet{golub1999molecular} this
distinction is critical for successful treatment which substantially
differs between classes. In fact, although remissions can be achieved
using any of these therapies, cure rates are markedly
increased and unwarranted toxicities are avoided when targeting the
specific type of leukemia with the right therapy.

\begin{figure}[!tpb]
  \centerline{\includegraphics[width=1\columnwidth]{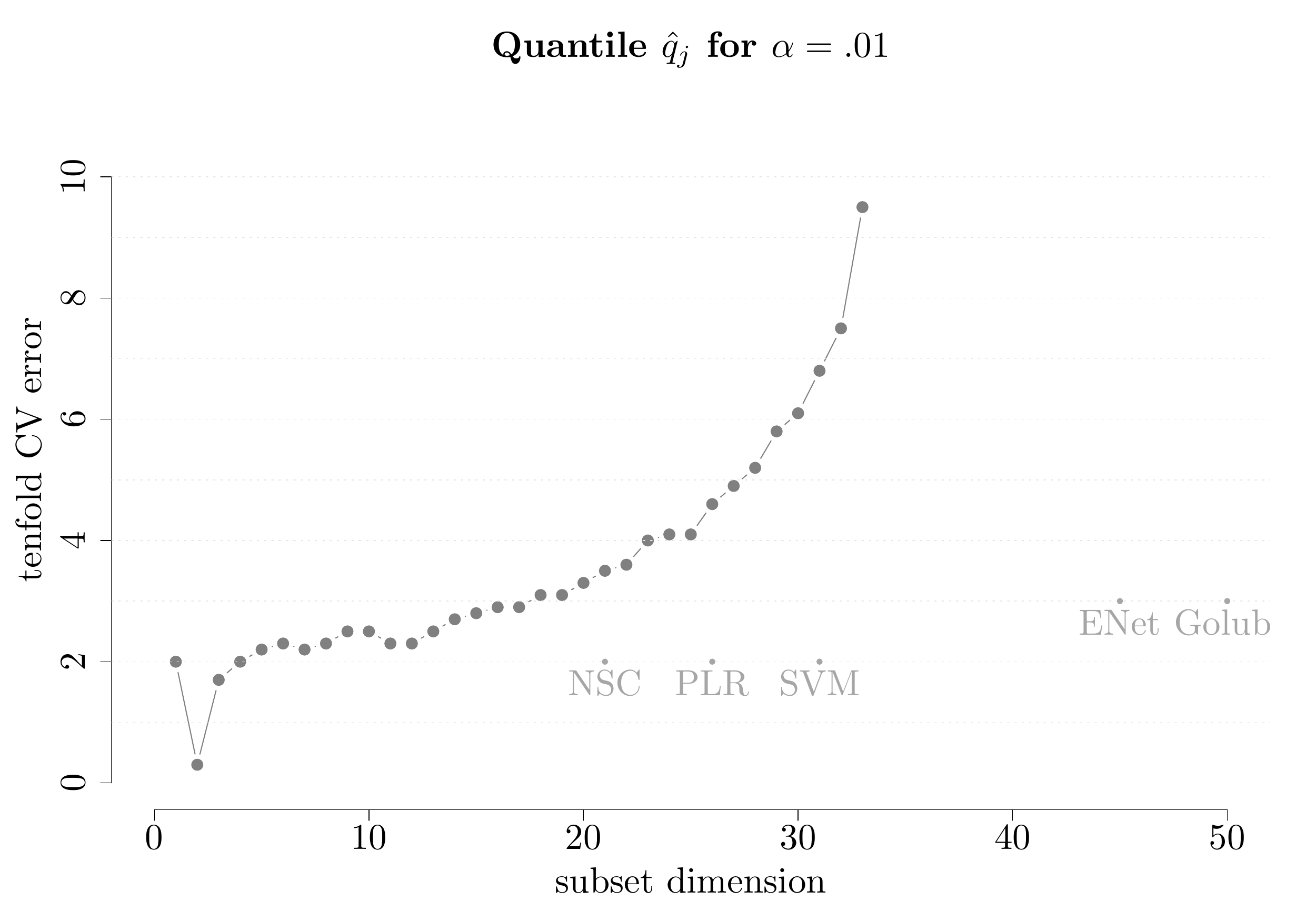}}
  \caption{Number of covariates vs. $\widehat{D}$ on leukemia cancer
    classification training set. The names are abbreviations for other
    selection method referred in Table \ref{Results}.} 
  \label{fig:curve}
\end{figure}

In order to understand how our proposed methodology performs
compared to existing ones, we split the
\textit{leukemia} data into the same training set (38
patients) and test set (34 patients) as in the original
work by \citet{golub1999molecular}. 
We employ our method on the training set to understand the dimension of the model
and to select the most relevant genes. Setting
$\alpha = 0.01$, the corresponding observed quantile
of the 10-fold cross-validation CE ($\widehat{D}$) is shown in Figure
\ref{fig:curve}. It can be seen that the error immediately decreases to
almost zero when using two covariates instead of one, after which
it monotonically increases, suggesting that the optimal model dimension
is two.
  
In Figure \ref{fig:curve} we also plotted the performance of the
other selection methods used on this training data which are
represented by labelled dots reporting the acronyms of these methods
that are listed in Table \ref{Results}. The approach proposed in this work compares favourably to all other methods in terms of prediction power. Indeed,
all the other methods lie under the curve to the right of its minimum
indicating that, compared to our method, they select models of
considerably higher dimensions without achieving the same degree of
performance in terms of CE. 
Therefore, for this particular case, our
method outperforms the other methods.
The sparsity and tenfold CV error are further illustrated 
in Table \ref{Results}, where we also present the average prediction error
on the test data. Considering the latter, it can be seen how the performance of the different methods are similar but the proposed method is able to achieve the same performance by selecting models of a considerably lower dimension.
As a final note to the table, the last line
reports the performance of model averaging. Indeed if the interest
lies in predicting, as described earlier, the algorithm of Section
\ref{methods} provides a set of models whose CE is below a given
quantile $\alpha$. The predictions of these models can be used in the
spirit of model averaging where a general prediction can be obtained
by taking the average of predictions of the selected set of
models. The proposed methodology can therefore be potentially seen as
a bridge between model selection and model averaging.

\begin{center}
\begin{table}[!ht]
		\centering
		\scalebox{.75}
		{
        \begin{tabular}{@{}lcccccc@{}}\toprule
        \textbf{Method}&\phantom{a}&\textbf{Tenfold CV} &\phantom{a}&\textbf{Test error}&\phantom{a}&\textbf{Number of} \\
         &&\textbf{error}&& && \textbf{genes} \\
        \midrule
         Golub && 3/38 && 4/34 && 50 \\
         Support vector machine  && 2/38 && 1/34 && 31 \\
         (with recursive feature elimination) &&  &&  &&  \\
         Penalised logistic regression && 2/38 && 1/34 && 26 \\
         (with recursive feature elimination) &&  &&  &&  \\
         Nearest shrunken centroids && 2/38 && 2/34 && 21 \\
         Elastic net && 3/38 && 0/34 && 45 \\[.1in]
         Proposed &&  &&  &&  \\ 
         Model a && 0/38 && 2/34 && 2 \\
         Model b && 0/38 && 2/34 && 2 \\
         Model c && 0/38 && 2/34 && 2 \\
         $[\ldots]$ &&  &&  &&  \\
         Model averaging &&  && 2/34 && 2 \\
         \hline
        \bottomrule
        \end{tabular}
        }        
        \caption{Summary of Leukemia classification results. The table
          is taken from \citet{zou2005regularization} except for the
          Proposed part. We obtained a total of 107 models of size 2
          (109 different biomarker) using a probability $\alpha=0.01$,
          $B=20000$ bootstrap replicates, a selection probability
          $\pi=.5$ with $D$ the tenfold-CV repeated 10 times. Model a
          to c are three examples out of the 107 models. All 107
          models have a tenfold-CV error of 0. The best test error is
          2 and the worst is 12. For model averaging all models are
          equally weighted. \label{Results}} 
\end{table}
\end{center}

Once this procedure is completed, we can create a gene
network to facilitate interpretation. This is a direct benefit of our method which does not deliver a single model after the selection process but provides a
series of models that can be linked to each other and interpreted jointly. Indeed, the
existence of a single model that links the covariates to the explained
variable is probably not realistic in many settings, especially for
gene classification. For this reason, the frequency with which each
gene is included within the selected models and with which these genes
are coupled with other genes provides the building block to create an
easy-to-interpret gene network with powerful explanatory and
predictive capacities. A graphical representation of this gene network can be found in Appendix \ref{sec:golub} together with a table where the biomarkers are listed according to their position
in the model. These positions represent families
of biomarkers (or genes) whose members are interchangeable. By the latter we mean that, given the presence of  biomarkers from other
families, specific biomarkers can be replaced by another biomarker from within
the same family without losing predictive power. 
This is the idea
behind finding a paradigmatic network for gene selection purposes. In the following paragraph we provide a summary biological interpretation of the the three main biomarkers (i.e. the most frequent in the selected models) which we call ``hubs'' from which the networks start.


\subsubsection{Biological interpretation}
The three hubs that were identified are the following:
\begin{enumerate}
	\item Cystatin C: a secreted cysteine protease inhibitor
          abundantly expressed in body fluids
          \citep[see][]{xu2015cystatin}; 
	\item Zyxin: a zinc-binding phosphoprotein that concentrates
          at focal adhesions and along the actin cytoskeleton; 
	\item Complement factor D: a rate-limiting enzyme in the
          alternative pathway of complement activation
          \citep[see][]{white1992human}. 
\end{enumerate}
In the current state of knowledge about acute leukemia, these three
hubs appear to make sense from a biological viewpoint. Cystatin C is
directly linked to many pathologic processes through various
mechanisms and recent studies indicate that the roles of Cystatin C in
neuronal cell apoptosis induction include decreasing B-cell leukemia-2
(BCL-2) whose deregulation is known to be implicated in resistant AML
\citep[see][]{sakamoto2015targeting}. Zyxin is a protein that
interacts with Vasodilator-stimulated phosphoprotein (VASP) with both
being involved in cellular adhesion and motility. VASP interacts with
ABL (breakpoint cluster region-abelson) and is a substrate of the
Bcr–Abl oncoprotein which drives oncogenesis in patients with chronic
myeloid leukemia (CML) due to a constitutive activation of tyrosine
kinase activity \citep[see][]{bernusso2015imatinib}. Further results
suggest that the phosphorylation and dephosphorylation cycle of VASP
by the Abi-1-bridged mechanism regulates association of VASP with
focal adhesions, which may regulate adhesion of Bcr-Abl-transformed
leukaemic cells \citep[see][]{masahiro2012abi}. Finally, Complement
factor D, together with several other components of both the classical
and alternative complement cascade, is primarily expressed through
both adipocytes and monocytes-macrophages in human subjects
\citep[see][]{white1992human, gabrielsson2003high}. A recent review in
\citet{ratajczak2014novel} has stressed the role of the complement
cascade as a trigger for hematopoietic stem cells from bone marrow
into blood. 
 
The interpretation of the network can be carried out through plots or tables
such as those presented in Appendix \ref{sec:golub} where the biomarkers can be grouped together into clusters having the same
biological traits, e.g. transcription/translation factor activity, DNA
repair and catabolism, apoptotic activity. This grouping allows a more straightforward
interpretation of the links between the different families thereby 
providing a more general overview of how the elements of the
identified network interact.

\section{Conclusions}
\label{conclusion}

This paper has proposed a new model selection method with various
advantages compared to existing approaches. Firstly, it
allows the user to specify the criterion according to which they would
like to assess the quality of a model. In this setting, it gives an
estimate of the dimension of the problem,
 allowing the user to
understand how many gene expressions are needed in a model to
well describe and predict the response of interest.
 Building on this, it provides a paradigmatic
structure of the selected models where the selected covariates are
considered as elements in an interconnected biological network. 
The approach can handle more
variables than observations without going through dimension-reduction
techniques such as pre-screening or penalization. 

The problem definition of
this method and the algorithmic structure used to solve it deliver
further advantages such as the ability to cope with noisy inputs,
missing data, multicollinearity and the capacity to deal with outliers
within the response and the explanatory variables (robustness). 

Some issues which must be taken into account concerning the proposed method are (i) its computational demand and (ii)
its need for an external validation. As far as the first aspect goes,
this can be considered indeed negligible compared to the time often
required to collect the data it should analyse and can be greatly
reduced according to the needs and requirements of the
user. Concerning the second aspect, external validation is a crucial
point which is often overlooked and is required for any model
selection procedure. In this sense, the proposed method does not differ from any other existing approach in terms of additional requirements. 

Having proposed a method with considerable advantages for 
gene selection using statistical ideas in model selection and machine
learning, we expect future research concerning the
statistical properties of this approach aiming at understanding its
asymptotic behaviour and developing inference tools, which will be
highly challenging and rewarding.


\section*{Acknowledgements}
We are very thankful to John Ramey (\url{http://ramhiser.com/}) for having processed the breast cancer and leukemia data set in Github and for having kindly answered our requests. \\
We thank Maria-Pia Victoria-Feser (Research Center for Statistics, University of Geneva, Switzerland) for her valuable comments and inputs as well as her institutional support. 

\section*{Funding and Conflict of interest} This research is supported
by the National
Science Foundation (DMS-1206693). No conflict of interest can be declared.


\FloatBarrier
\bibliographystyle{natbib}
\bibliography{bibioinfo}

\begin{thebibliography}{}

\bibitem[Andres and Wittliff(2012)Andres and Wittliff]{andres2012co}
Andres, S.~A. and Wittliff, J.~L. (2012).
\newblock Co-expression of genes with estrogen receptor-$\alpha$ and
  progesterone receptor in human breast carcinoma tissue.
\newblock {\em Hormone molecular biology and clinical investigation\/}, {\bf
  12}(1), 377--390.

\bibitem[Bernusso {\em et~al.}(2015)Bernusso, Machado-Neto, Pericole, Vieira,
  Duarte, Traina, Hansen, Saad, and Barcellos]{bernusso2015imatinib}
Bernusso, V.~A., Machado-Neto, J.~A., Pericole, F.~V., Vieira, K.~P., Duarte,
  A.~S., Traina, F., Hansen, M.~D., Saad, S. T.~O., and Barcellos, K.~S.
  (2015).
\newblock Imatinib restores vasp activity and its interaction with zyxin in
  bcr--abl leukemic cells.
\newblock {\em Biochimica et Biophysica Acta (BBA)-Molecular Cell Research\/},
  {\bf 1853}(2), 388--395.

\bibitem[Bohrer {\em et~al.}(2014)Bohrer, Chuntova, Bade, Beadnell, Leon,
  Brady, Ryu, Goldberg, Schmechel, Koopmeiners, {\em
  et~al.}]{bohrer2014activation}
Bohrer, L.~R., Chuntova, P., Bade, L.~K., Beadnell, T.~C., Leon, R.~P., Brady,
  N.~J., Ryu, Y., Goldberg, J.~E., Schmechel, S.~C., Koopmeiners, J.~S., {\em
  et~al.} (2014).
\newblock Activation of the fgfr--stat3 pathway in breast cancer cells induces
  a hyaluronan-rich microenvironment that licenses tumor formation.
\newblock {\em Cancer research\/}, {\bf 74}(1), 374--386.

\bibitem[Cantoni {\em et~al.}(2007)Cantoni, Field, Mills~Flemming, and
  Ronchetti]{cantonietal2007}
Cantoni, E., Field, C., Mills~Flemming, J., and Ronchetti, E. (2007).
\newblock Longitudinal variable selection by cross-validation in the case of
  many covariates.
\newblock {\em Statistics in medicine\/}, {\bf 26}(4), 919--930.

\bibitem[Chin {\em et~al.}(2006)Chin, DeVries, Fridlyand, Spellman,
  Roydasgupta, Kuo, Lapuk, Neve, Qian, Ryder, {\em et~al.}]{chin2006genomic}
Chin, K., DeVries, S., Fridlyand, J., Spellman, P.~T., Roydasgupta, R., Kuo,
  W.-L., Lapuk, A., Neve, R.~M., Qian, Z., Ryder, T., {\em et~al.} (2006).
\newblock Genomic and transcriptional aberrations linked to breast cancer
  pathophysiologies.
\newblock {\em Cancer cell\/}, {\bf 10}(6), 529--541.

\bibitem[Chou {\em et~al.}(2010)Chou, Provot, and Werb]{chou2010gata3}
Chou, J., Provot, S., and Werb, Z. (2010).
\newblock Gata3 in development and cancer differentiation: cells gata have it!
\newblock {\em Journal of cellular physiology\/}, {\bf 222}(1), 42--49.

\bibitem[Christer {\em et~al.}(2013)Christer, Peter, Margaret, Stephen, and
  Kathryn]{christer2013mechanism}
Christer, H., Peter, K., Margaret, L.~A., Stephen, H., and Kathryn, M.~T.
  (2013).
\newblock A mechanism for epithelial-mesenchymal transition and anoikis
  resistance in breast cancer triggered by zinc channel zip6 and stat3 (signal
  transducer and activator of transcription 3).
\newblock {\em Biochemical Journal\/}, {\bf 455}(2), 229--237.

\bibitem[Chung {\em et~al.}(2014)Chung, Giehl, Wu, and Vadgama]{chung2014stat3}
Chung, S.~S., Giehl, N., Wu, Y., and Vadgama, J.~V. (2014).
\newblock Stat3 activation in her2-overexpressing breast cancer promotes
  epithelial-mesenchymal transition and cancer stem cell traits.
\newblock {\em International journal of oncology\/}, {\bf 44}(2), 403--411.

\bibitem[D{\'\i}az-Uriarte and De~Andres(2006)D{\'\i}az-Uriarte and
  De~Andres]{diaz2006gene}
D{\'\i}az-Uriarte, R. and De~Andres, S.~A. (2006).
\newblock {Gene Selection and Classification of Microarray Data using Random
  Forest}.
\newblock {\em BMC Bioinformatics\/}, {\bf 7}(1), 3.

\bibitem[Dudoit {\em et~al.}(2002)Dudoit, Fridlyand, and
  Speed]{dudoit2002comparison}
Dudoit, S., Fridlyand, J., and Speed, T.~P. (2002).
\newblock {Comparison of Discrimination Methods for the Classification of
  Tumors using Gene Expression Data}.
\newblock {\em Journal of the American statistical association\/}, {\bf
  97}(457), 77--87.

\bibitem[Gabrielsson {\em et~al.}(2003)Gabrielsson, Johansson, L{\"o}nn,
  Jern{\aa}s, Olbers, Peltonen, Larsson, L{\"o}nn, Sj{\"o}str{\"o}m, Carlsson,
  {\em et~al.}]{gabrielsson2003high}
Gabrielsson, B.~G., Johansson, J.~M., L{\"o}nn, M., Jern{\aa}s, M., Olbers, T.,
  Peltonen, M., Larsson, I., L{\"o}nn, L., Sj{\"o}str{\"o}m, L., Carlsson, B.,
  {\em et~al.} (2003).
\newblock High expression of complement components in omental adipose tissue in
  obese men.
\newblock {\em Obesity research\/}, {\bf 11}(6), 699--708.

\bibitem[George and McCulloch(1993)George and
  McCulloch]{georgeandmcculloch1993}
George, E. and McCulloch, R. (1993).
\newblock Variable selection via gibbs sampling.
\newblock {\em Journal of the American Statistical Association\/}, {\bf
  88}(423), 881--889.

\bibitem[George and McCulloch(1997)George and
  McCulloch]{georgeandmcculloch1997}
George, E.~I. and McCulloch, R.~E. (1997).
\newblock Approaches for bayesian variable selection.
\newblock {\em Statistica sinica\/}, {\bf 7}(2), 339--373.

\bibitem[Golub {\em et~al.}(1999)Golub, Slonim, Tamayo, Huard, Gaasenbeek,
  Mesirov, Coller, Loh, Downing, and Caligiuri]{golub1999molecular}
Golub, T.~R., Slonim, D.~K., Tamayo, P., Huard, C., Gaasenbeek, M., Mesirov,
  J.~P., Coller, H., Loh, M.~L., Downing, J.~R., and Caligiuri, M.~A. (1999).
\newblock {Molecular Classification of Cancer: Class Discovery and Class
  Prediction by Gene Expression Monitoring}.
\newblock {\em Science\/}, {\bf 286}(5439), 531--537.

\bibitem[Juang {\em et~al.}(1997)Juang, Hou, and Lee]{juangetal1997}
Juang, B., Hou, W., and Lee, C. (1997).
\newblock Minimum classification error rate methods for speech recognition.
\newblock {\em Speech and Audio Processing, IEEE Transactions on\/}, {\bf
  5}(3), 257--265.

\bibitem[Kouros-Mehr {\em et~al.}(2006)Kouros-Mehr, Slorach, Sternlicht, and
  Werb]{kouros2006gata}
Kouros-Mehr, H., Slorach, E.~M., Sternlicht, M.~D., and Werb, Z. (2006).
\newblock Gata-3 maintains the differentiation of the luminal cell fate in the
  mammary gland.
\newblock {\em Cell\/}, {\bf 127}(5), 1041--1055.

\bibitem[Kristensen {\em et~al.}(2012)Kristensen, Vaske, Ursini-Siegel,
  Van~Loo, Nordgard, Sachidanandam, S{\o}rlie, W{\"a}rnberg, Haakensen,
  Helland, {\em et~al.}]{kristensen2012integrated}
Kristensen, V.~N., Vaske, C.~J., Ursini-Siegel, J., Van~Loo, P., Nordgard,
  S.~H., Sachidanandam, R., S{\o}rlie, T., W{\"a}rnberg, F., Haakensen, V.~D.,
  Helland, {\AA}., {\em et~al.} (2012).
\newblock Integrated molecular profiles of invasive breast tumors and ductal
  carcinoma in situ (dcis) reveal differential vascular and interleukin
  signaling.
\newblock {\em Proceedings of the National Academy of Sciences\/}, {\bf
  109}(8), 2802--2807.

\bibitem[Masahiro {\em et~al.}(2012)Masahiro, Mizuho, Yunfeng, Masayoshi,
  Ryosuke, Takuya, Norihiro, Tatsuo, and Naoki]{masahiro2012abi}
Masahiro, M., Mizuho, S., Yunfeng, Y., Masayoshi, I., Ryosuke, F., Takuya, O.,
  Norihiro, I.-K., Tatsuo, T., and Naoki, W. (2012).
\newblock Abi-1-bridged tyrosine phosphorylation of vasp by abelson kinase
  impairs association of vasp to focal adhesions and regulates leukaemic cell
  adhesion.
\newblock {\em Biochemical Journal\/}, {\bf 441}(3), 889--899.

\bibitem[Ratajczak(2014)Ratajczak]{ratajczak2014novel}
Ratajczak, M. (2014).
\newblock A novel view of the adult bone marrow stem cell hierarchy and stem
  cell trafficking.
\newblock {\em Leukemia\/}.

\bibitem[Sakamoto {\em et~al.}(2015)Sakamoto, Grant, Saleiro, Crispino, Hijiya,
  Giles, Platanias, and Eklund]{sakamoto2015targeting}
Sakamoto, K.~M., Grant, S., Saleiro, D., Crispino, J.~D., Hijiya, N., Giles,
  F., Platanias, L., and Eklund, E.~A. (2015).
\newblock Targeting novel signaling pathways for resistant acute myeloid
  leukemia.
\newblock {\em Molecular genetics and metabolism\/}, {\bf 114}(3), 397--402.

\bibitem[Taniguchi and Karin(2014)Taniguchi and Karin]{taniguchi20146}
Taniguchi, K. and Karin, M. (2014).
\newblock Il-6 and related cytokines as the critical lynchpins between
  inflammation and cancer.
\newblock In {\em Seminars in immunology\/}, volume~26, pages 54--74. Elsevier.

\bibitem[Tibshirani {\em et~al.}(2002)Tibshirani, Hastie, Narasimhan, and
  Chu]{tibshirani2002diagnosis}
Tibshirani, R., Hastie, T., Narasimhan, B., and Chu, G. (2002).
\newblock {Diagnosis of Multiple Cancer Types by Shrunken Centroids of Gene
  Expression}.
\newblock {\em Proceedings of the National Academy of Sciences\/}, {\bf
  99}(10), 6567--6572.

\bibitem[Vaske {\em et~al.}(2010)Vaske, Benz, Sanborn, Earl, Szeto, Zhu,
  Haussler, and Stuart]{vaskeetal2010}
Vaske, C.~J., Benz, S.~C., Sanborn, J.~Z., Earl, D., Szeto, C., Zhu, J.,
  Haussler, D., and Stuart, J.~M. (2010).
\newblock Inference of patient-specific pathway activities from
  multi-dimensional cancer genomics data using {PARADIGM}.
\newblock {\em Bioinformatics\/}, {\bf 26}(12), i237--i245.

\bibitem[White {\em et~al.}(1992)White, Damm, Hancock, Rosen, Lowell, Usher,
  Flier, and Spiegelman]{white1992human}
White, R.~T., Damm, D., Hancock, N., Rosen, B., Lowell, B., Usher, P., Flier,
  J., and Spiegelman, B. (1992).
\newblock Human adipsin is identical to complement factor d and is expressed at
  high levels in adipose tissue.
\newblock {\em Journal of Biological Chemistry\/}, {\bf 267}(13), 9210--9213.

\bibitem[Xu {\em et~al.}(2015)Xu, Ding, Li, and Wu]{xu2015cystatin}
Xu, Y., Ding, Y., Li, X., and Wu, X. (2015).
\newblock Cystatin c is a disease-associated protein subject to multiple
  regulation.
\newblock {\em Immunology and cell biology\/}.

\bibitem[Yang and Song(2010)Yang and Song]{yangandsong2010}
Yang, A.-J. and Song, X.-Y. (2010).
\newblock Bayesian variable selection for disease classification using gene
  expression data.
\newblock {\em Bioinformatics\/}, {\bf 26}(2), 215--222.

\bibitem[Zhang(2011)Zhang]{zhang2011}
Zhang, T. (2011).
\newblock Adaptive forward-backward greedy algorithm for learning sparse
  representations.
\newblock {\em Information Theory, IEEE Transactions on\/}, {\bf 57}(7),
  4689--4708.

\bibitem[Zhu and Hastie(2004)Zhu and Hastie]{zhu2004classification}
Zhu, J. and Hastie, T. (2004).
\newblock {Classification of Gene Microarrays by Penalized Logistic
  Regression}.
\newblock {\em Biostatistics\/}, {\bf 5}(3), 427--443.

\bibitem[Zou and Hastie(2005)Zou and Hastie]{zou2005regularization}
Zou, H. and Hastie, T. (2005).
\newblock {Regularization and Variable Selection via the Elastic Net}.
\newblock {\em Journal of the Royal Statistical Society: Series B\/}, {\bf
  67}(2), 301--320.

\end{thebibliography}

\newpage
\appendix

\section{Example for Model (\ref{eq:model})} 
\label{sec:example}

A few examples of model (\ref{eq:model}) are as follows: 
\begin{itemize}
    \item $Y$ can be a continuous variable representing, for example,
      the concentration of a certain substance in the body. In this
      case, a common model is $Y=\X\trans\bb^{\jmath}+\epsilon$, which
      corresponds to $g(\X,\bb^{\jmath})=\X\trans\bb^{\jmath}$. 
    \item $Y$ can also be a count variable, such as when it represents
      the number of tumours in a patient. A common Poisson model is 
    \begin{equation*}
    	\mathbb{P}(Y=y\mid \X)=\exp(-e^{\X\trans\bb^{\jmath}})e^{\X\trans\bb^{\jmath}y}/y!,
    \end{equation*}
which can be equivalently written as
$Y=e^{\X\trans\bb^{\jmath}}+\epsilon$, where
$\epsilon=\exp(-e^{\X\trans\bb^{\jmath}})e^{\X\trans\bb^{\jmath}Y}/Y!-e^{\X\trans\bb^{\jmath}}$
has mean zero and $g(\X,\bb^{\jmath})=e^{\X\trans\bb^{\jmath}}$. 
    \item $Y$ can be a Bernoulli random variable ($Y \in \{0,1\}$),
      when it records the presence/absence of a tumour. A common
      logistic model is 
\begin{equation*}
\mathbb{P}(Y=1\mid\X)=1-1/(e^{\X\trans\bb^{\jmath}}+1)
\end{equation*}
and can be equivalently written as
$Y=1-1/(e^{\X\trans\bb^{\jmath}}+1)+\epsilon$, where
$\epsilon=Y-1+1/(e^{\X\trans\bb^{\jmath}}+1)$ also has mean zero. Thus
$g(\X,\bb^{\jmath})=1-1/(e^{\X\trans\bb^{\jmath}}+1)$. 
\end{itemize}

\section{Adapting the algorithm to $p$}
\label{sec:ada}

In this subsection we provide two variants of the algorithm proposed in Section \ref{methods} in order to adapt it to situations where $p$ is either small or large. 

\subsection{Adapting the algorithm to very large $p$}

In situations where $p$ is extremely large and the initial step of the algorithm is not computationally feasible, this step can, for example, be replaced by the following modified initial step:

\begin{enumerate}
	\item[$\text{A}'$.] \emph{Large $p$ Modified Initial Step:} We start by augmenting our initial variable set $\mathcal{M}_0$ with $d = 1$ variable in order to construct the set $\mathcal{I}^*_1$.
	\begin{enumerate}
		\item[1.] Augment $\mathcal{M}_0$ with $d = 1$ variable selected uniformly at random in $\mathcal{J}_f$.
		\item[2.] Construct $B$ models of dimension $1$ by repeating Step $\text{A}'$.1 $B$ times.
		\item[3.] From Steps $\text{A}'$.1 and $\text{A}'$.2, construct the set $\mathcal{I}^*_1$ using (\ref{eq:formsetI}). Go to Step B and let $d = 2$.
	\end{enumerate}
\end{enumerate}

\subsection{Adapting the algorithm to small $p$} 

On the other hand, when $p$ is of reasonable size it may be possible
to compute and evaluate all the $\binom{p}{d'}$ models of dimension
$2\le d' \leq d_{\text{max}}$. In such cases, it may be feasible to
also modify the initial step of the proposed algorithm to a different
modified initial step. A possible modification is the following:

\begin{enumerate}
	\item[$\text{A}''$.] \emph{Small $p$ Modified Initial Step:} We start
          by augmenting our initial varable set $\mathcal{M}_0$ with
          $d$ ($1 \leq d \leq d'$) variables in order to construct the
          sets $\mathcal{I}^*_{1}, ..., \mathcal{I}^*_{d'}$. 
	\begin{enumerate}
		\item[1.] We augment our initial variable set
                  $\mathcal{M}_0$ with 1 variable in order to
                  construct the set $\mathcal{I}^*_1$. 
		\begin{enumerate}
			\item[(i)] Construct the $p$ possible models obtained by augmenting 
$\mathcal{M}_0$ with each of the $p$ available variables.

			\item[(ii)] Compute $\widehat{D}(\cdot,\cdot)$
                          for every model obtained in Step (i). 
			\item[(iii)] From Steps (i) and (ii),
                          construct the set $\mathcal{I}^*_1$  using
                          (\ref{eq:formsetI}). Go to Step $\text{A}''$.2 and let
                          $d = 2$. 
		\end{enumerate} 

		\item[2.] We augment our initial model $\mathcal{M}_0$
                  set by $d$ variables  in order to construct the set
                  $\mathcal{I}^*_d$. 
			\begin{enumerate}
				\item[(i)] Construct the
                                  $\binom{p}{d}$ possible models and
                                  augment $\mathcal{M}_0$ with all
                                  variables of these constructed
                                  models. 
				\item[(ii)] Compute $\widehat{D}$ for every model obtained in Step (i).
				\item[(iii)] From Steps (i) and (ii),
                                  construct the set $\mathcal{I}^*_d$
                                  using (\ref{eq:formsetI}) and let $d
                                  = d+1$. Go to Step $\text{A}''$.2 (if $d< d'$)
                                  or Step B.1 (if $d\ge d'$), with model dimension starting value $d$.
			\end{enumerate}
	\end{enumerate}
\end{enumerate}

\section{Determining the Models’ Dimension via Testing}
\label{sec:test} 

The type of test and its corresponding rejection level are determined
by the user based on the nature of the divergence measure. For
example, if we take the $L_1$ loss function as a divergence, one could
opt for the Mann-Whitney test or if the loss function is a
classification error (as in the applications in Section
\ref{results}), one could choose the binomial test or other tests for
proportions.  
The rejection level will depend, among others, on the number of tests
that need to be run, typically less than $d_{\text{max}}-1$, and need
to be adjusted using, for example, the Bonferroni correction. 
Finally, once the set $\mathcal{S}^*_{d^*}$ is obtained, the user may
still want to ``filter'' the resulting models.  
Indeed, the number of models in the solution $\mathcal{S}^*_{d^*}$ may
be large and the corresponding divergence estimates may vary
considerably from model to model. Since these divergence measures are
estimators, we again propose a multiple testing procedure to reduce
the number of models in $\mathcal{S}^*_{d^*}$. Before doing so, we
eliminate redundant models, thereby making sure that every model is
included only once. Then, we start the testing procedure with an empty
set $\mathcal{S}^0_{d^*} = \emptyset$ to which we add the model (or
one of the models) that has the minimum divergence measure estimate,
denoted $\wh D_{\jmath_{\text{min}}}$, where
$\jmath_{\text{min}}\in\mathcal{S}^*_{d^*}$ denotes this model. Then
for every model
$\jmath\in\mathcal{S}^*_{d^*}\backslash\jmath_{\text{min}}$, we test
whether $\wh D_{\jmath}$ is greater than $\wh D_{\jmath_{\text{min}}}$.
We add the model to $\mathcal{S}^0_{d^*}$ if the difference is not
significant and stop adding models as soon as the test deems that the
divergence of the next model is indeed larger. 
 By doing so we finally obtain
 $\mathcal{S}^0_{d^*}\subseteq\mathcal{S}^*_{d^*}$ which is the set
 containing the models (and hence covariates) which can be interpreted
 in a paradigmatic network.


\section{Related literature}
\label{sec:related}

Some of the ideas put forth in this work have also been considered
in the literature. An extensive survey of the related works goes
beyond the scope of this paper. Here we briefly describe some of the
connections to three main ideas that have been explored previously.  

The first one is recognizing that practitioners might aim to minimize
some criterion that differs from likelihood-type losses. An
interesting paper illustrating this point is \cite{juangetal1997} in
the context of speech recognition. For their classification problem,
these authors propose to minimize a ``smoothed'' version of the
decision rule used for classification. The advantage of this procedure
is that it yields better misclassification errors than using pure
likelihood based criteria which intrinsically fit a distribution to
the data. In the approach presented in this work we also
  deliver an approximate solution but, as opposed to approximating the
  problem and solving the latter in an exact manner as in
  \cite{juangetal1997}, we define the exact problem and try to
  approximately minimize the misclassification error through our
  algorithm.

Second, there is a large literature that uses stochastic search
procedures to explore the space of candidate models. Influential work
in this direction include \cite{georgeandmcculloch1993} and
\cite{georgeandmcculloch1997} who postulate hierarchical Bayesian
models. In their set up, subsets of promising predictors form models
with higher posterior probabilities. An interesting application of
this framework for disease classification using gene expression data
is the work of \cite{yangandsong2010}. \cite{cantonietal2007} also
consider a random exploration of the space of possible models, but
avoiding the Bayesian formulation of
\cite{georgeandmcculloch1993}. Their approach defines a probability
distribution for the various candidate models based on a
cross-validated prediction error criterion and then uses a Markov
Chain Monte-Carlo method to generate a sample from this probability
distribution. 
An important feature of the stochastic search implied by our algorithm
is that it is a greedy method, while the aforementioned methods are
not.  
The typical forward/backward greedy algorithms proposed in the
literature are not random, while existing stochastic procedures are
not greedy. 
{\color{black}Thus, the combination of greedy approach and random search
  approach seems to be new.} See for instance \cite{zhang2011} for
some theory on greedy algorithms in sparse scenarios.

Third, other authors have also considered providing a set of
interesting models as opposed to a single ``best'' model.  The
stochastic search procedures mentioned in the above paragraph can
naturally be used to obtain a group of interesting models. For
example, \cite{cantonietal2007} consider a set of best
indistinguishable models in terms of prediction. Random forests can be
used to select variables and account for the stability of the chosen
model as in \cite{diaz2006gene}. These methods can also be used to
construct a set of interesting models. 

\section{More results on Acute Leukemia}
\label{sec:golub}

\begin{figure}[!tpb]
  \centerline{\includegraphics[width=0.8\columnwidth]{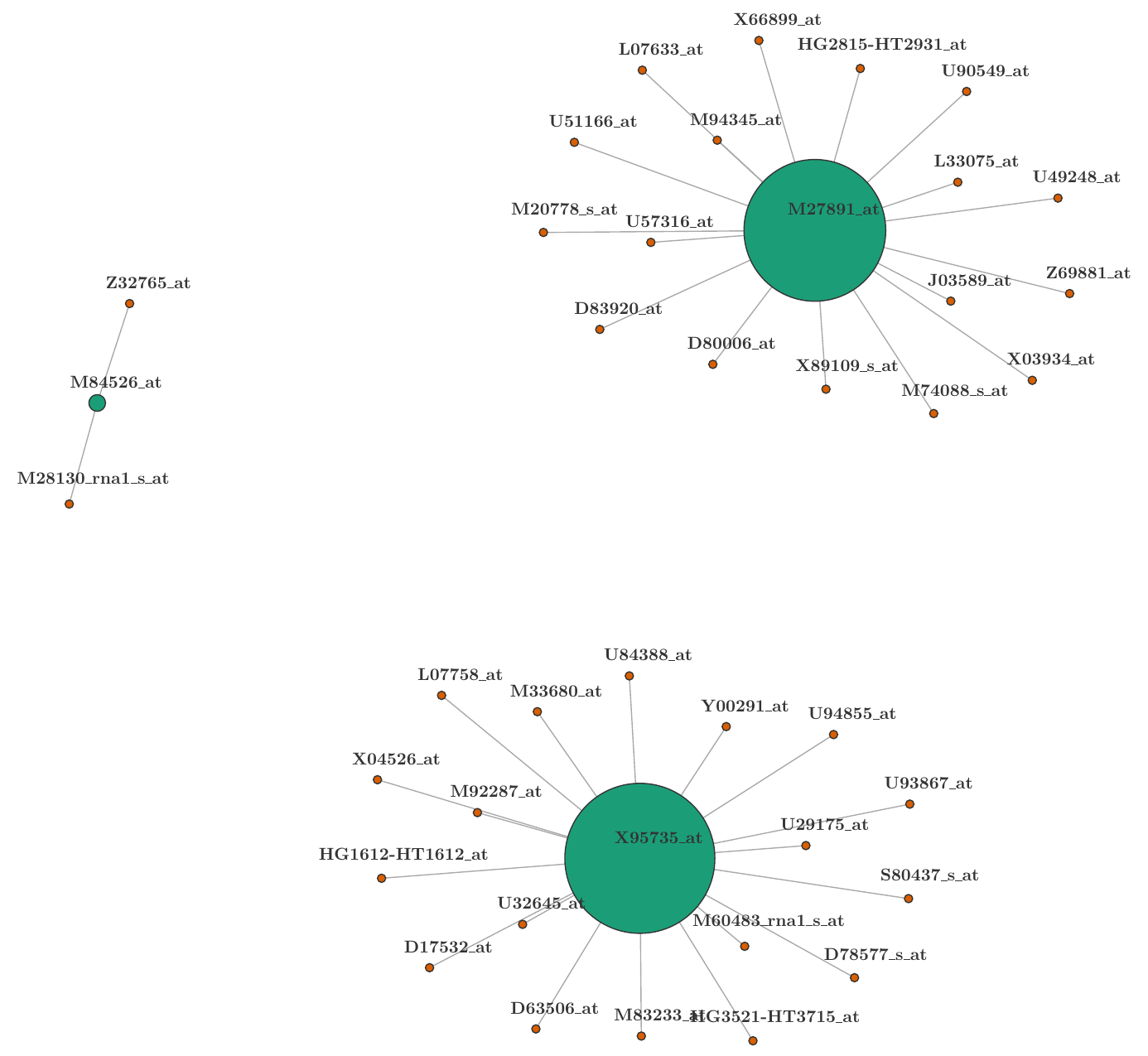}}
  \caption{Network representation of biomarkers selected from
    \textit{leukemia} data-set. Colors represent the position of
    covariates within the model: green for first position (hub) and
    orange for second. 
    The width of the
    connecting lines is proportional to the frequency with which two
    biomarkers appear in the same model. The size of the disk is
    proportional to the frequency with which a biomarker is present
    within the selected set of models.}
  \label{fig:network:golub}
\end{figure}

Figure \ref{fig:network:golub} graphically represents the network created by the proposed method for the \textit{leukemia} data-set
where the size of a disk represents the frequency with which a
particular biomarker is included in the selected models, and the
line connecting the disks indicates the biomarkers that are included in the same
model. Since the model dimension in this case is two, each biomarker
is connected with only one other biomarker and, as can be observed, the
proposed method identifies three main ``hubs'' for the networks (green
disks) generating three networks.

Table \ref{Leuk} reports the main biomarker hubs and related biomarker networks for the \textit{leukemia} data set analysed in Section \ref{golub}. 

\begin{table*}[!ht]
		\centering
		\scalebox{.6}
		{
        \begin{tabular}{lllll}\toprule
        &\textbf{Affy ID} & \textbf{Gene ID} & \textbf{Gene Function} & \textbf{Biological} \\
		&	&	& & \textbf{Process}\\
        \midrule
       \multicolumn{2}{l}{\textbf{NETWORK 1}}& \\
        Position 1& M27891{\_}at	& 	 ENSG00000101439 & Cystatin C & AA\\[.1in]
        Position 2 & D80006{\_}at & ENSG00000114978 &	MOB kinase activator 1A & AA	\\
        &  M20778{\_}s{\_}at&  ENSG00000163359	& Collagen, type VI, alpha 3 & AA	 \\
        & U57316{\_}at   & ENSG00000108773	& K(lysine) acetyltransferase 2A & TF \\
         & U90549{\_}at &  ENSG00000182952	 & High mobility group nucleosomal binding domain 4 & TF	 \\
                  & X66899{\_}at & ENSG00000182944 &	Ewing Sarcoma region 1; RNA binding protein	 	 & TF \\
            &  M74088{\_}s{\_}at& ENSG00000134982  &	Adenomatous polyposis coli, DP2, DP3, PPP1R46 & TF \\
            &  U51166{\_}at&  ENSG00000139372	 &	thymine-DNA glycosylase & TF \\
         & Z69881{\_}at&   ENSG00000074370 &	ATPase, Ca++ transporting, ubiquitous	& IPT \\
            &  U49248{\_}at&  ENSG00000023839	 	& ATP-binding cassette, sub-family C (CFTR/MRP), member 2 & IPT	\\
         &  X89109{\_}s{\_}at&  ENSG00000102879	 	&	Coronin, actin binding protein, 1A & IPT\\
         &  HG2815-HT2931{\_}at&   ENSG00000092841	 &	Myosin, Light Chain, Alkali, Smooth Muscle (Gb:U02629) & ACC\\
        & M94345{\_}at &   ENSG00000042493	&	 Capping protein (actin filament), gelsolin-like & ACC	 \\
        & L33075{\_}at &  ENSG00000140575	&	 IQ motif containing GTPase activating protein 1 & ACC	 \\
         &  L07633{\_}at& ENSG00000092010  	 &	Proteasome (prosome, macropain) activator subunit 1 (PA28 alpha) & APC\\
        &  J03589{\_}at&  ENSG00000102178	&	Ubiquitin-like 4A & APC \\
         
         &  D83920{\_}at&  ENSG00000085265	 &	FCN1, Ficolin-1 & IR \\
         &  X03934{\_}at&  ENSG00000167286	&	CD3d molecule, delta (CD3-TCR complex) & IR	 \\
[.2in]
         \hline
        \multicolumn{2}{l}{\textbf{NETWORK 2}}& \\
         Position 1& X95735{\_}at	& 	 ENSG00000159840 &  Zyxin	& ACC \\[.1in]
         Position 2 & X04526{\_}at  &	ENSG00000185838 & Guanine nucleotide binding protein (G protein), beta polypeptide 1 & ST	 \\ 
        &  D78577{\_}s{\_}at&   ENSG00000128245 	&	 Tyrosine 3-monooxygenase/tryptophan 5-monooxygenase & ST	 \\
        &                   &                       & activation protein, eta & \\
         & U32645{\_}at&  ENSG00000102034	 &	E74-like factor 4 (ets domain transcription factor) & TF	 \\
         & U93867{\_}at &  ENSG00000186141	&	Polymerase (RNA) III (DNA directed) polypeptide C (62kD) & TF	 \\
        & U29175{\_}at &  ENSG00000127616	 &	\parbox[t]{10cm}{SWI/SNF related, matrix associated, actin dependent regulator of chromatin, subfamily A, member 4} & TF \\
        &  Y00291{\_}at& ENSG00000077092 	&	 Retinoic acid receptor, beta & TF	 \\
         &  D17532{\_}at &  ENSG00000110367	 &	 DEAD (Asp-Glu-Ala-Asp) Box Helicase 6 & TF \\ 
         & HG3521-HT3715{\_}at & ENSG00000127314 &	Ras-Related Protein Rap1b & TF \\
        &  M83233{\_}at&  ENSG00000140262	&	 Transcription factor 12 & TF \\
         & U94855{\_}at &  ENSG00000175390	&	Eukaryotic translation initiation factor 3, subunit F & TF \\
        &  L07758{\_}at&  ENSG00000136045	 &	PWP1 homolog & TF \\
         &  D63506{\_}at&  ENSG00000116266	& Syntaxin binding protein 3 & IR \\
        &  M33680{\_}at&   ENSG00000110651	 &	CD81 molecule &	IR \\ 
         &  HG1612-HT1612{\_}at&  ENSG00000175130  & Macmarcks & CG	 	 \\
         &  M92287{\_}at&  ENSG00000112576	&	Cyclin D3 & CG	 \\
         &  M60483{\_}rna1{\_}s{\_}at&  ENSG00000113575	 	& Protein Phosphatase 2 (formerly 2A), catalytic subunit, alpha isoform	& CG \\
         &  U84388{\_}at& ENSG00000169372 	&	CASP2 and RIPK1 domain containing adaptor with death domain & AA \\
        &  S80437{\_}s{\_}at&  ENSG00000169710	 	&	Fatty acid synthase  \\[.2in]
         \hline
         \multicolumn{2}{l}{\textbf{NETWORK 3}}& \\
         Position 1 & M84526{\_}at	& 	 ENSG00000197766  	& Complement factor D (adipsin) & IR\\[.1in]
         Position 2 & M28130{\_}rna1{\_}s{\_}at  & ENSG00000169429 & Interleukine-8 & IR	 \\
         &Z32765{\_}at &  ENSG00000135218 & CD36 - Thrombospondin receptor	& IR	 \\

         \bottomrule
        \end{tabular} 
        }        
        \caption{Biomarker network organisation - \emph{leukemia} data set - Lymphoblastic / Myeloblastic leukemia. TF = Transcription/translation factor activity, DNA repair and catabolism -  AA = apoptotic activity -  IR = immunity, inflammatory response (blood coagulation, antigen presentation and  complement activation) -  IPT = intracellular protein trafficking, transmembrane transport -  ACC = actin activity, cytoskeleton organisation -  APC = protein catabolism - ST = intracelular signal transduction - CG = cell growth, proliferation and division. Source: \url{www.ensembl.org};  \url{www.uniprot.org}} \label{Leuk}
\end{table*}

\section{Breast Cancer}
\label{sec:breast} 
 
\begin{figure}[!tpb]
  \centerline{\includegraphics[width=0.8\columnwidth]{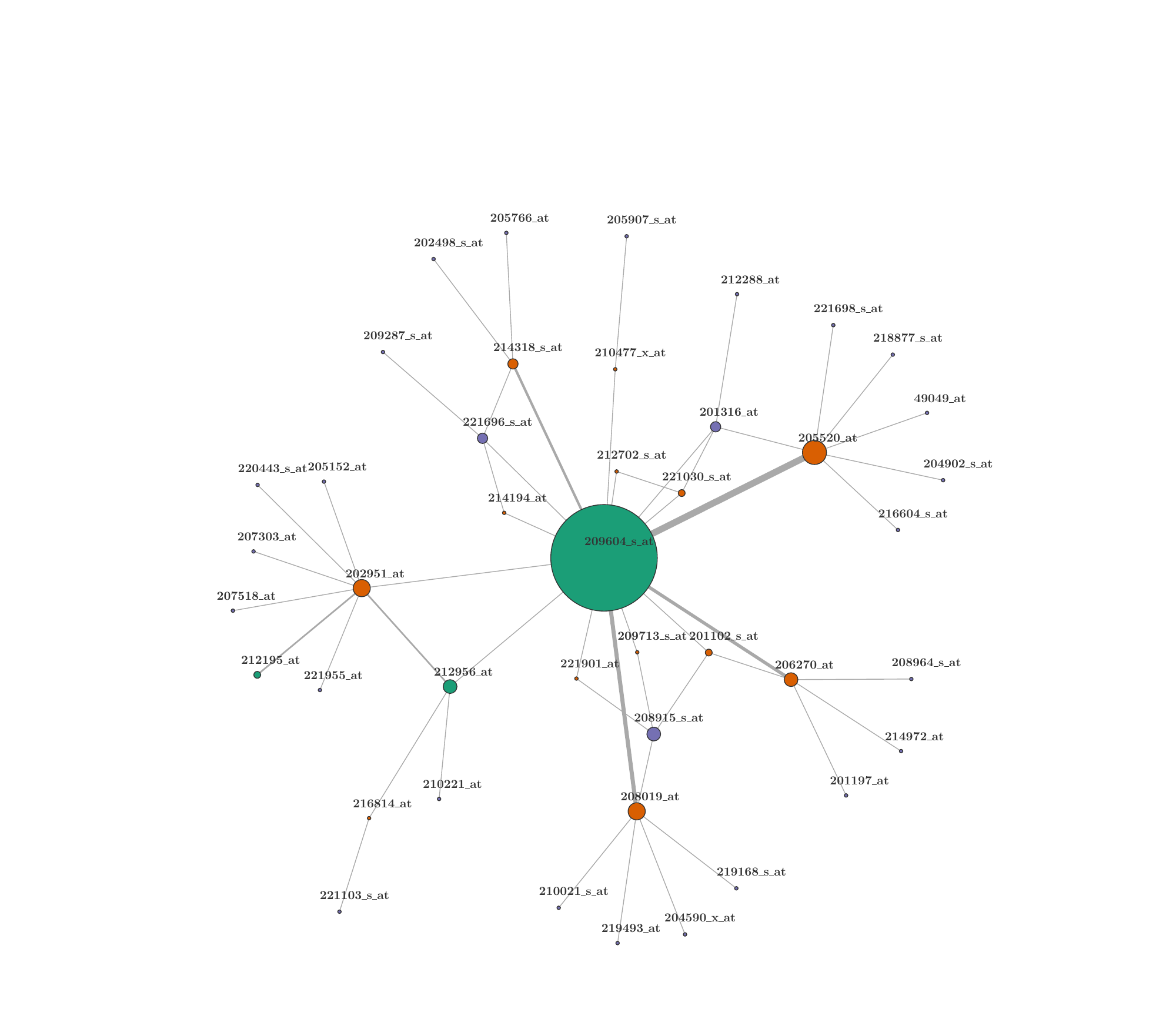}}
  \caption{Network representation of biomarkers selected from
    \textit{breast cancer} data-set. Colors represent the position of
    covariates within the model: green for first position (hub),
    orange for second and purple for third. The width of the
    connecting lines is proportional to the frequency with which two
    biomarkers appear in the same model. The size of the circles is
    proportional to the frequency with which a biomarker is present
    within the selected set of models. (Note: biomarker
    ``209602{\_}s{\_}at'' is merged with biomarker
    ``209604{\_}s{\_}at'').} 
  \label{fig:network:chin}
\end{figure}

The second data-set we analyzde is the \textit{breast cancer} data
presented in \citet{chin2006genomic} for which we only provide a
summary biological interpretation of the results of the proposed
method, having already discussed the working of our approach in
Section \ref{golub}. The main goal behind analyzing this data is to
identify the estrogen receptor expression on tumor cells which is a
crucial step for the correct management of breast cancer. Figure  
\ref{fig:network:chin} shows the paradigmatic network identified by
our method for the \textit{breast cancer} data for which the selected
model dimension is three (i.e. only three biomarkers are needed in a
model to well classify the breast cancer). Table \ref{Chin} provides
the details of the networks based on the three main hubs and is to be
interpreted as described in Section \ref{golub}. 
 
This figure is a clear example of the advantages of the proposed
method since, it not only  selects a  set of low-dimensional models
with a high predictive power, but also provides the basis for a more
general biological interpretation which takes into account
interactions between different biomarkers as opposed to one single
model. 
The three main hubs identified through the proposed algorithm are:
\begin{enumerate}
	\item GATA binding protein 3 (GATA3): a transcription factor
          regulating the differentiation of breast luminal epithelial
          cells; 
	\item IL6 Signal Transducer (IL6 ST): a pro-inflammatory cytokine signal transducer;
	\item TBC1 domain family, member 9 (TBC1D9): a
          GTPase-activating protein for Rab family protein involved in
          the expression of the ER in breast tumors. 
\end{enumerate}
 GATA3 is known to regulate the differentiation of epithelial cells in
 mammary glands \citep[see][]{kouros2006gata} and is required for
 luminal epithelial cell differentiation. Its expression is
 progressively lost during luminal breast cancer progression as cancer
 cells acquire a stem cell-like phenotype
 \citep[see][]{chou2010gata3}. IL6 ST has been linked to breast cancer
 epithelial-mesenchymal transition and cancer stem cell traits
 \citep[see][]{chung2014stat3}, cancer-promoting microenvironment
 \citep[see][]{bohrer2014activation} and resistance
 \citep[see][]{christer2013mechanism}. Moreover, this result supports
 the assertion by \citet{taniguchi20146} that IL6 ST and related
 cytokines are the critical lynchpins between inflammation and
 cancer. Finally, concerning the third biomarker, a recent publication
 by \citet{andres2012co} has shown that the expression of the ER on
 the surface of breast tumor cells is highly correlated with the
 coordinate expression of different genes among which we can find
 TBC1D9 and GATA3. These two genes are not only considered as relevant
 genes according to the proposed method but as actual hubs of the
 ``best'' models which define the structure of the identified
 network.
 Instead of selecting a single model with many biomarkers whose
 interactions may be difficult to interpret, the proposed method
 selects a set of models with few biomarkers that allow them to be
 individually easy to interpret without losing the possibility of
 interpreting them within the larger network. This is what this paper
 intends with the expression ``paradigmatic network'' since by taking
 this approach it is possible to identify a  set of biomarker families
 within which each biomarker is interchangeable with the others. 

\bigskip

\newpage 

		{\tiny
        \begin{longtable}{@{}p{.14\textwidth}p{.12\textwidth}p{.18\textwidth}p{.34\textwidth}p{.105\textwidth}@{}}
        \toprule
        & \textbf{Affy ID} & \textbf{Gene ID}& \textbf{Gene Function}& \textbf{Biological Process}\\
        \midrule
        \textbf{NETWORK 1} \\
        Position 1 & 209604{\_}s{\_}at	& ENSG00000107485	  &GATA binding protein 3 & {\color{black} TF}\\[.1in]
        Position 2 & 205520{\_}at   &  ENSG00000115808
	& Striatin, calmodulin binding protein& ER   \\
        Position 3 &204902{\_}s{\_}at & ENSG00000168397  & Autophagy related 4B, cysteine peptidase (APG4B, AUTL1, DKFZp586D1822, KIAA0943)&APC \\
         & 221698{\_}s{\_}at & ENSG00000172243   & C-type lectin domain family 7, member A&IR \\
         & 49049{\_}at &  ENSG00000178498  &  Deltex 3, E3 ubiquitin ligase &APC \\
         & 209602{\_}s{\_}at &   ENSG00000107485  & GATA binding protein 3& TF\\
         & 216604{\_}s{\_}at &  ENSG00000003989  & Solute carrier family 7 (cationic amino acid transporter, y+ system), member 2 &IPT
        \\
         & 218877{\_}s{\_}at &   ENSG00000066651  &  TRNA methyltransferase 11 homolog & TF \\
         & 201316{\_}at &  ENSG00000106588 & Proteasome (prosome, macropain) subunit, alpha type, 2&APC \\
         \\
        Position 2      &  208019{\_}at &  ENSG00000147117  	 &	Zinc finger protein 157& TF \\
        Position 3     &  219168{\_}s{\_}at & ENSG00000186654 & PRR5 (Proline rich 5 (renal))&CG		 \\
        &  219493{\_}at &  ENSG00000171241 & SHC SH2-domain binding protein 1	&CG	 \\
        &  204590{\_}x{\_}at & ENSG00000139719 & Vacuolar protein sorting 33 homolog A	&APC	 \\
        &  210021{\_}s{\_}at &  ENSG00000152669 & 	Cyclin O&CG \\
        & 208915{\_}s{\_}at & ENSG00000103365 & 	Golgi-associated, gamma adaptin ear containing, ARF binding protein 2&IPT	 \\
        \\
        Position 2 & 214318{\_}s{\_}at     &   ENSG00000073910    &  Furry homolog&ACC   \\
    Position 3 &  205766{\_}at  &   ENSG00000173991    & Titin-cap (Telethonin) &ACC    \\
    &  221696{\_}s{\_}at & ENSG00000060140   & Serine/threonine/tyrosine kinase 1&CG		 \\
    & 202498{\_}s{\_}at & ENSG00000059804  & Solute carrier family 2 (facilitated glucose transporter), member 3&STM		 \\
    \\
    Position 2 & 201102{\_}s{\_}at   &  ENSG00000141959    & Phosphofructokinase, liver&STM    \\
    Position 3 & 208915{\_}s{\_}at   &  ENSG00000103365    & Golgi-associated, gamma adaptin ear containing, ARF binding protein 2&IPT    \\
    \\
    Position 2 & 201316{\_}at & ENSG00000106588 & Proteasome (prosome, macropain) subunit, alpha type, 2 &APC \\
    Position 3 &  212288{\_}at   &   ENSG00000187239   &  Formin binding protein 1  &ACC  \\
    \\
    Position 2 & 209713{\_}s{\_}at   &  ENSG00000116704  & Solute carrier family 35 (UDP-GlcA/UDP-GalNAc     transporter), member D1   &STM  \\
    Position 3 & 208915 {\_}s{\_}at   &   ENSG00000103365   &  Golgi-associated, gamma adaptin ear containing, ARF binding protein 2 &IPT \\
   \\
    
   Position 2 & 212702{\_}s{\_}at   &  ENSG00000185963   & Bicaudal D homolog 2 &ACC    \\
    Position 3 &  221030{\_}s{\_}at  &  ENSG00000138639    &   Rho GTPase activating protein 24 &ACC  \\
   \\
   Position 2 & 212956{\_}at   & ENSG00000109436  &  TBC1 domain family, member 9 (with GRAM domain)&IPT \\
    Position 3 &  210221{\_}at  &      ENSG00000080644 &  Cholinergic receptor, nicotinic, alpha 3 (neuronal) &ITT  \\
   \\
   Position 2 & 214194{\_}at   &  ENSG00000083520  &  DIS3 mitotic control homolog (Ribosomal RNA-processing protein 44) & TF  \\
    Position 3 & 221696{\_}s{\_}at    &   ENSG00000060140  &Serine/threonine/tyrosine kinase 1 &CG  \\
\\
   Position 2 & 216814{\_}at &    ENSG00000232267   & ACTR3 pseudogene 2   &PUP \\
   Position 3 & 221103{\_}s{\_}at   &  ENSG00000206530    &  Cilia and flagella associated protein 44&ACC \\
   \\
   Position 2 & 221030{\_}s{\_}at   & ENSG00000138639    &   Rho GTPase activating protein 24 &ACC  \\
    Position 3 & 201316{\_}at    &   ENSG00000106588   & Proteasome (prosome, macropain) subunit, alpha type, 2  &APC  \\
   \\
   Position 2 & 221696{\_}s{\_}at   &  ENSG00000060140   &  Serine/threonine/tyrosine kinase 1&CG   \\
    Position 3 &  209287{\_}s{\_}at   &  ENSG00000070831    &  Cell division control protein 42 homolog &ACC    \\
   \\
   Position 2 & 221901{\_}at   & ENSG00000138944   &    KIAA1644&PUP  \\
    Position 3 & 208915{\_}s{\_}at   &   ENSG00000103365   &  Golgi-associated, gamma adaptin ear containing, ARF binding protein 2&IPT    \\
    \\
    Position 1 & 209602{\_}s{\_}at & ENSG00000107485	  &GATA3& TF\\
    \\
    Position 2 & 202951{\_}at   &    ENSG00000112079   & Serine/threonine kinase 38  &CG  \\
    Position 3 & 220443{\_}s{\_}at   &   ENSG00000116035   &  VAX2 (ventral anterior homeobox 2) & TF  \\
    &  221955{\_}at & ENSG00000088256 & Guanine nucleotide binding protein (G protein), alpha 11 (Gq class)&ITT		 \\
     & 207303{\_}at & ENSG00000154678 & 	Phosphodiesterase 1C, calmodulin-dependent 70kDa	&ST \\
     &  205152{\_}at & ENSG00000157103 & Solute carrier family 6, member 1	&ST	 \\
       &  207518{\_}at & ENSG00000153933 & Diacylglycerol kinase, epsilon 64kDa&ST 		 \\
    \\
    Position 2 &  206270{\_}at     &   ENSG00000126583    &   Protein kinase C, gamma&ST  \\
    Position 3 &  208964{\_}s{\_}at  &   ENSG00000149485    &    Fatty acid desaturase 1 &FAM\\
    &  201197{\_}at & ENSG00000123505 & 	Adenosylmethionine decarboxylase 1&CG	 \\
    &  201102{\_}s{\_}at & ENSG00000141959 & ATP-dependent 6-phosphofructokinase, liver type	&STM	 \\
    &  214972{\_}at &  ENSG00000198408  &  Protein O-GlcNAcase (Meningioma expressed antigen 5 (hyaluronidase))&ST	 \\
    \\
    Position 2 & 210477{\_}x{\_}at   &  ENSG00000107643  &  Mitogen-activated protein kinase 8 &CG  \\
    Position 3  &  205907{\_}s{\_}at & ENSG00000127083  & Osteomodulin&STM	 \\
   \\[.2in]
 \hline
            \textbf{NETWORK 2} \\
            Position 1 &212195{\_}at	& ENSG00000134352	  &  IL6 Signal Transducer	& ICT \\[.1in]
         Position 2 & 202951{\_}at  &	ENSG00000112079 & Serine/threonine kinase 38 & CG	 \\ 
         Position 3 & 221955{\_}at & ENSG00000088256 & Guanine nucleotide binding protein (G protein), alpha 11 (Gq class) & ITT \\
         & 207303{\_}at & ENSG00000154678 & Phosphodiesterase 1C, calmodulin-dependent 70kDa& ICT \\
          \\ [.2in]
         \hline 
         \textbf{NETWORK 3} \\
         Position 1 & 212956{\_}at	& 	ENSG00000109436  	& TBC1 domain family, member 9 (with GRAM domain) & IPT \\[.1in]
         Position 2 & 202951{\_}at &  ENSG00000112079 & Serine/threonine kinase 38&CG	 \\
 Position 3  & 205152{\_}at&   ENSG00000157103 & Solute carrier family 6, member 1 &ST	 \\ 
             & 207518{\_}at &     ENSG00000153933  &  Diacylglycerol kinase, epsilon 64kDa&ST\\
             \\
 Position 2 &   216814{\_}at    &  ENSG00000232267       &  ACTR3 pseudogene 2 &PUP   \\
 Position 3 & 221103{\_}s{\_}at   & ENSG00000206530      &    Cilia and flagella associated protein 44 &ACC   \\
\bottomrule
\caption{Biomarker network organisation - \emph{breast cancer} data set - Estrogen Receptor - Breast Cancer. \\TF = Transcription/translation factor activity, DNA/RNA  repair and catabolism -  ER = estrogen receptor activity -   APC = autophagy - protein catabolism -  IR =  immunity, inflammatory response (blood coagulation, antigen presentation and  complement activation)  -   CC = cell/cell communication -  ST =  intracellular signal transduction, protein glycosylation -   CG = cell growth and division - IPT = intracellular protein trafficking , transmembrane amino-acid transporter - ACC = actin activity, cytoskeleton organisation, cell projection - STM = sugar transport and metabolism -  ITT = ion transmembrane transport, transmembrane signaling systems - PUP = pseudogene, uncharacterized protein - FAM =  fatty acid metabolism. Source: \url{www.uniprot.org}; \url{www.ncbi.nlm.nih.gov/gene}
} \label{Chin}
\end{longtable}}

\end{document}